\documentclass[11pt]{article} 
\usepackage[dvips]{epsfig}  

\newcommand{\be}{\begin{equation}}
\newcommand{\ee}{\end{equation}}
\newcommand{\bea}{\begin{eqnarray}}
\newcommand{\eea}{\end{eqnarray}}

\begin{document}


\vspace{1cm}

\begin{center}
{\large\bf One-dimensional extended Bose-Hubbard model with a confining potential: a DMRG analysis}
\end{center}
\vspace{5mm}

\begin{center}

{\large Laura Urba$^{\scriptstyle 1}$, Emil Lundh$^{\scriptstyle 2, 3}$, Anders Rosengren$^{\scriptstyle 4}$. \\ }

\vspace{5mm} 
Condensed matter theory, Department of theoretical physics, KTH, AlbaNova University Center, SE-106 91 Stockholm, Sweden.

\vspace{5mm}
July 2006

\end{center}

\vspace{6mm}

\begin{center}
{\large \bf Abstract}
\end{center}
\vspace{1cm}

The extended Bose-Hubbard model in a quadratic trap potential is
studied using a finite-size density-matrix renormalization group
method (DMRG). We compute the boson density profiles, the local
compressibility and the hopping correlation functions.  We observed
the phase separation induced by the trap in all the quantities studied
and conclude that the local density approximation is valid in the
extended Bose-Hubbard model. From the plateaus obtained in the local
compressibility it was possible to obtain the phase diagram of the
homogeneous system which is in agreement with previous results.
\noindent

\vspace{2cm}
\begin{flushleft}
PACS \hspace{0.2cm} 03.75.Nt~~~05.30.Jp~~~71.45.Lr~~~72.80.Sk~~~73.43.Nq  
\end{flushleft}
\vfill
\begin{flushleft}
{\tt
$^1$ lgurba@kth.se \\
$^2$ emil.lundh@tp.umu.se \\
$^3$ Present address: Department of Physics, Ume{\aa} University, 
SE-90187 Ume{\aa}.\\
$^4$ roseng@kth.se
}
\end{flushleft}


\section{Introduction}
Recently, much attention has been paid to the study of cold atoms in optical lattices 
\cite{jaksch,greiner}. The absence of disorder and the low temperatures at which these experiments are 
performed make them excellent systems in which to study quantum phase transitions. 
Theoretically, the simplest Hamiltonian that models a system of bosonic atoms in an optical lattice 
is the Bose-Hubbard (BH) model \cite{jaksch}. This Hamiltonian has two parameters: a hopping kinetic energy 
$t$ and an on-site repulsion $U$, whose ratio $t/U$ is widely tunable since it 
depends exponentially on the strength of the optical lattice potential. 
At zero temperature and for small $t/U$, the BH model presents two phases: a 
superfluid (SF) and a Mott-insulator (MI), the last one appearing when the number of bosons is a multiple of 
the number of sites. At large $t/U$ only the superfluid phase (SF) is present \cite{fisher}. 
In the experiments with optical lattices, however, an external trap is applied in order to confine the atoms and this means 
that different phases can appear in different, spatially separated, parts of 
the system. From a theoretical point of view, an 
inhomogeneous term representing the confining potential has to be added 
to the Hamiltonian. The resulting inhomogeneous 
BH model has been studied in one dimension (1D) by several authors using different 
approaches \cite{batrouni,kollath,pollet,emil,sara}.   
It has been concluded that the phase diagram of the 1D inhomogeneous system can be obtained 
with good accuracy using a local-density approximation \cite{sara}. In 2D and 3D, the BH model in a 
trap has been studied using quantum Monte Carlo simulations in \cite{kashu,wessel} among other works.  

Recent progress with cooling and condensing Chromium atoms 
\cite{chromium} suggests that an extension to the BH model is experimentally 
relevant. 
In contrast to the alkali atoms, which have been used in all 
optical-lattice experiments so far, $^{52}$Cr atoms have a large dipole moment 
which leads to long-range, dipolar interactions. Such interactions are 
proportional to the inverse cube of the atomic separation and are anisotropic, 
so that by confining the atoms to one dimension and polarizing the dipoles, 
one may tune the ratio between the short-range 
and dipole parts of the interaction terms to any desired value by tuning the 
polarization angle \cite{goral}. This makes $^{52}$Cr atoms in an 
optical lattice a potentially ideal system for experimentally exploring the 
phase diagram of the Bose-Hubbard model with long-range interactions. 

 To understand the effect of longer-range repulsive interactions among the atoms a term can be added to the Hamiltonian that 
takes into account the interaction between nearest neighbors. The obtained Hamiltonian is the so called Extended 
Bose-Hubbard (EBH) model \cite{kuhner,kovrizhin}.
 In this work we study the EBH model in a quadratic confining potential using a finite-size 
density-matrix renormalization group method (finite-size DMRG) \cite{white}. The DMRG is a numerical 
method which gives very accurate results for the ground state properties of one-dimensional interacting systems. In our 
case we compute the boson density profiles, its variance (or local compressibility) and the hopping correlation 
functions and we show the validity of the local-density approximation. This allows us to recover the phase diagram of the
homogeneous EBH model by studying the system in a trap. 

 Other recent work on this model is \cite{hebert} where Monte Carlo simulations are used to study solitons that 
appear when the half-filled system is doped. 
    
This paper is organized as follows: in Sec.~2 we give a brief description of the EBH model and its 
phases, in Sec.~3  we discuss the numerical method, in Sec.~4 we present our results and finally, in Sec.~5 
we give our conclusions.        
\section{The Extended Bose-Hubbard model}
The Hamiltonian corresponding to the one-dimensional  EBH model is
\be
H_{EBH}= -t \sum_i (b^{\dagger}_i b_{i+1} +  b_i b^{\dagger}_{i+1}) + U/2 \sum_i n_i (n_i-1) + V \sum_i n_i n_{i+1}-\mu 
\sum_i n_i ,
\label{EBHHam}
\ee
where $b^{\dagger}_i$ ($b_i$) creates (annihilates) a boson in the site $i$, $n_i= b^{\dagger}_i b_i$ is the boson 
number operator and $t$, $U$, $V$ and $\mu$ are assumed to be 
positive. The term proportional to $t$ takes into account the hopping of bosons between nearest-neighbor sites 
(kinetic energy), the second term is an on-site repulsion proportional to $U$, the third term corresponds to a 
repulsion of strength $V$ between nearest-neighbor bosons, and in the fourth term, $\mu$ is the chemical potential.   
In this paper we will set $U=1$, i.e. all the parameters will be measured in units of $U$. 

\begin{figure}
\centering
\epsfysize=7.0truecm
\epsffile{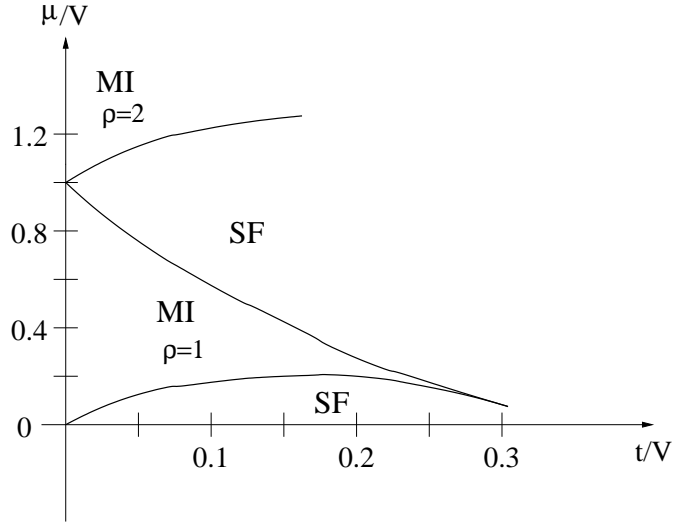}
\caption{The phase diagram of the Bose-Hubbard model showing the first Mott-insulator (MI) lobes at density 
$\rho=1$ and $\rho=2$ surrounded by the superfluid (SF) first published in Ref.~\cite{batrou}.}
\label{bhdia}
\end{figure}

\begin{figure}
\centering
\epsfysize=7.0truecm
\epsffile{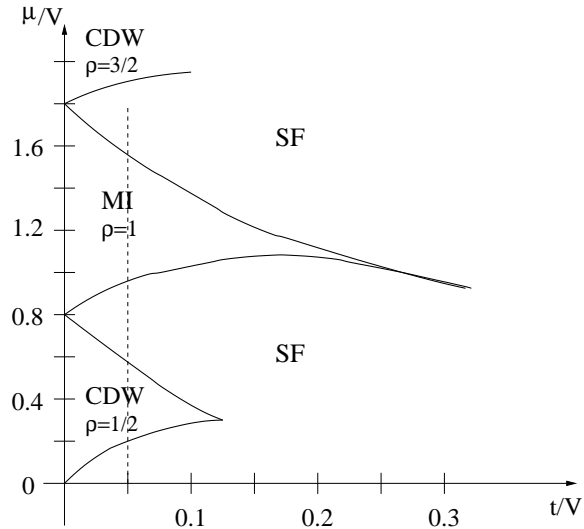}
\caption{The phase diagram of the extended Bose-Hubbard model with $V=0.4$ first published in Ref.~\cite{kuhner}. The first lobe corresponds to 
the charge density wave phase (CDW) at $\rho=1/2$ and the second to the Mott-insulator (MI) at density $\rho=1$. The dotted line 
indicates how a system in a spatially varying potential will scan the phase diagram, giving rise to phase separation.}
\label{ebhdia}
\end{figure}

If $V=0$, the Hamiltonian reduces to the ordinary 
BH Hamiltonian. As mentioned above, its phase diagram presents two 
phases for small $t/U$: a superfluid (SF) and a Mott-insulator (MI), as is shown in Fig.~\ref{bhdia} \cite{batrou}. At $U=0$ 
(or $t/U \rightarrow \infty$), the system is in the SF phase which is characterized by a non-zero superfluid density, 
correlation functions that decay with a power law behavior and a gap-less energy spectrum. As the tunneling decreases relative to the on-site repulsion, 
it becomes  more difficult to move the bosons along the system and, at a critical value of $t/U$, the system 
undergoes a phase transition to a Mott-insulator. The energy spectrum in the MI phase has a gap which implies that the 
correlation functions decay exponentially. 
The number of bosons per site is constant which means that the boson density is commensurate with the number of sites 
(see the different lobes in Fig.~\ref{bhdia}) and the compressibility is zero. 
For $V\neq 0$ a new phase, called the charge density wave (CDW) phase, develops at half integer densities and small $t/U$, giving rise to the phase diagram shown in Fig.~\ref{ebhdia} \cite{kuhner}. This CDW  phase is an insulator characterized by strong local density fluctuations.
Like the 
Mott-insulator, its excitation spectrum has a gap, the correlation functions decay exponentially and it is incompressible. 
In order to confine the atoms in an optical lattice, a magnetic trapping 
potential is usually applied in experiments. Such traps are 
faithfully modeled by adding a quadratic potential term to the Hamiltonian 
defined in Eq.\ (\ref{EBHHam}): 
\be
H_{trap}=\sum_{i=0}^L \frac{w^2}{2}(i-L/2)^2 n_i.
\label{trap}
\ee      
The potential is defined so that it is zero at the center of the chain. 
The potential strength parameter $w$ must in numerical simulations be 
chosen to be large enough to assure that there are no 
bosons at the edges of the chain. The addition of this potential 
makes the system inhomogeneous and as a consequence different phases may  
appear in different regions. This phase separation has been studied using exact diagonalization, Monte Carlo and DMRG 
techniques in the $V=0$ case, i.e.\ the conventional BH model 
\cite{batrouni,kollath,emil,sara}. On the other hand, the $V \neq 0$ case, 
i.e.\ the EBH model, has been 
investigated in the absence of a trap using a finite-size DMRG in 
Ref.~\cite{kuhner} and using a Gutzwiller ansatz in Ref.~\cite{kovrizhin}. 

The local density approximation will be valid if the local properties of the inhomogeneous system can be determined 
assuming that different sites in the lattice have different effective chemical potentials. The site 
dependent chemical potential $\mu_i$ is defined by subtracting the trap potential from the chemical potential $\mu$,  
\be
\mu_i = \mu - \frac{w^2}{2}(i-L/2)^2.
\ee   
If the local-density approximation works, then for a given set of parameters of the Hamiltonian, the local properties of the 
inhomogeneous system can be obtained from the phase diagram of the homogeneous system following a path 
along a vertical line with $t/U$=constant, as illustrated in Fig.~\ref{ebhdia}. 
\section{Numerical method} \label{nummeth}
We use a finite-size density-matrix renormalization group algorithm \cite{white} to determine the density profiles 
and the correlation functions of the extended Bose-Hubbard model in a quadratic potential. 
With this technique, the system is built up by successively adding sites 
up to a predetermined length and, after that, a series of sweeps are 
performed until the process converges. In this way, the optimal approximate ground state is obtained \cite{white2}. 
We use open boundary conditions and the usual superblock configuration, see
for instance Ref.~\cite{laura}. 

When studying an inhomogeneous system, it is usually necessary to perform a 
large number of sweeps in order to reach convergence.
In the present case, we find it useful to employ a mean field ansatz as a starting density profile in order 
to obtain a faster convergence and reduce the number of sweeps. In this way, the procedure of adding bosons while increasing 
the system size in the DMRG procedure is done in a more efficient way. 

 A small complication is that the mean field calculation is performed with a simple code where the chemical potential 
is fixed. In the DMRG code, however, it is the total number of bosons that is fixed, so we recalculate the chemical potential 
as $dE_0/dN_b$ (where $E_0$ is the ground state energy and $N_b$ is the total number of bosons) around the 
considered $N_b$. We explain this in detail in the following section. 

Dealing with a bosonic system means that the number of states per site is infinite. This implies an extra 
truncation in the DMRG related to the maximum number of bosons per site ($nb_s$) allowed. Following 
\cite{kuhner} we considered up to 4 bosons per site (5 states per site) in our results. For all the cases considered, 
the maximum value in the density profiles was less than 1.5 leaving the cut-off of 4 bosons per site well above this 
maximum. In fact, for other values  $nb_s \ge 2$, we did not observe 
differences in the resulting density profiles, which shows that our results 
are well converged in this respect.

The truncation error ($\Delta$), which is the sum of the density matrix eigenvalues discarded in each DMRG step, 
gives a measure of the numerical errors. It strongly depends on the system's correlation length \cite{kuhner}, 
increasing considerably with it. At a fixed number of states kept, we expect a smaller $\Delta$ in the insulating phase 
(where the correlation length is finite) than in the superfluid phase (where the correlation length is infinite).
Since in our case we have phase separation due to the trap, we find that the truncation 
error increases with the hopping $t$ because more superfluid phase is present in the system 
(see Fig.~\ref{ebhdia}), so we increased the number of states kept in the density-matrix as we increased $t$.            
We obtained good accuracy with a truncation error $\Delta < 10^{-7}$. 
\section{Results} \label{ressec}
\begin{figure}
\centering
\epsfysize=7.0truecm
\epsffile{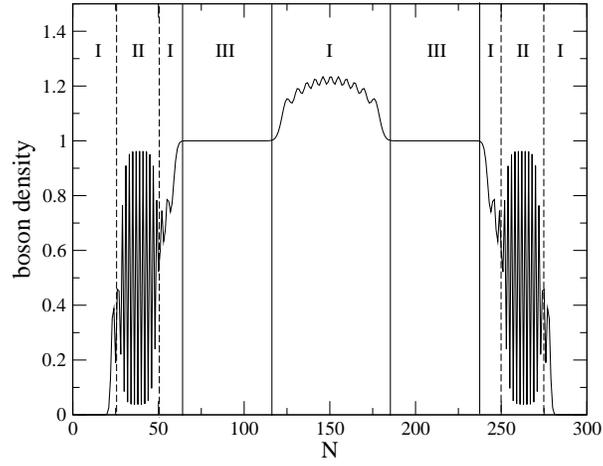}
\caption{Boson density vs N for $t=0.05$, $V=0.4$, and $w=0.015$. The total number of bosons is $N_{bos}=233$.}
\label{bosden1}
\end{figure}

To calculate the boson density profiles we evaluated $\langle n_i \rangle$ in the last sweep of the 
DMRG, where $n_i$ is the boson number operator and $\langle \, \rangle$ stands for the expectation value in the ground state. 
With the superblock configuration B-x-x-B (where B corresponds to a block and x to a site), the site $i$ is chosen 
as one of the sites x and is moved along the system as we perform the last sweep. This has the advantage that the matrices 
$n_i$ are more accurate. They are clearly truncated because we consider only up to 5 states 
per site, but they are not affected by the density matrix truncation that comes after rotating the block matrices in each DMRG 
step. An example of a density profile is shown in Fig.~\ref{bosden1} where the parameters in the Hamiltonian are $t=0.05$, $U=1$ 
and $V=0.4$, and the value of $w=0.015$ in the trap potential is large enough to ensure there are no bosons at the edges of the 
chain. The total number of bosons was $N_{bos}=233$ and the length of the chain $N=300$. As an ansatz for the DMRG we used a 
boson density profile obtained from a mean field calculation where we fixed the chemical potential at $\mu=1.8$; this gave 
us a profile to start the DMRG with a total of $233$ bosons.                

In the density plot shown in Fig.~\ref{bosden1}, the vertical lines separate the different phases located in 
the regions labeled as I, II and III. In the regions II we can see the fluctuating  
behavior of the local density that characterizes the charge density wave phase (CDW). The wavelength of the oscillations is $2$ (in units of the lattice spacing) and the local density fluctuates between values very close to $1$ and $0$. 
This is an improvement respect to previous results (see Fig. 4 in Ref.~\cite{hebert}) where the local density oscillations 
in the CDW regions are not so well established, specially in the case with the largest number of bosons.
  
\begin{figure}
\centering
\epsfysize=7.0truecm
\epsffile{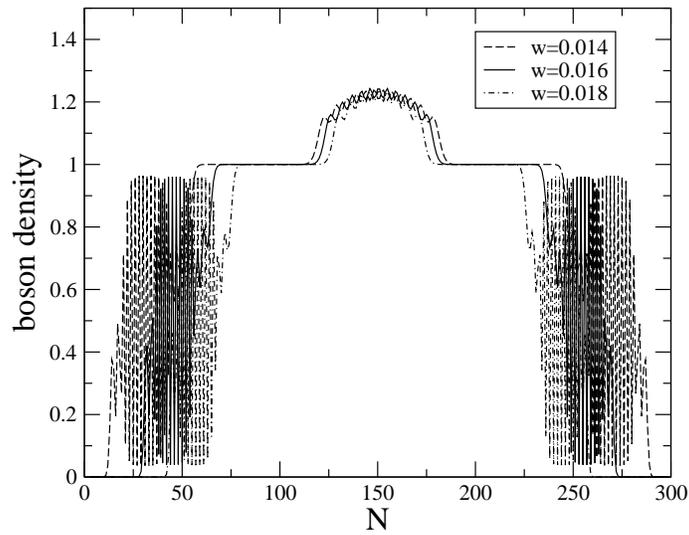}
\caption{Boson density vs N for $t=0.05$, $V=0.4$, and different values of $w$.}
\label{dendifw}
\end{figure}

The plateaus in the regions III clearly correspond to the Mott-insulator 
where there is only one boson per site (i.e. the local density equals 1). The regions labeled as I correspond 
to the superfluid phase. In particular, an almost quadratic dependence expected for the superfluid from the 
mean field approximation (MFA) can be observed at the center of the chain. The short wavelength oscillations on 
top of that are, as pointed out by K{\"u}hner {\it et al.} \cite{kuhner}, finite-size effects. 
In Ref.~\cite{kuhner} the authors were interested in the infinite homogeneous system and therefore argued 
that those oscillations should be disregarded because they decrease towards the center of the chain. In our case, however, the 
boundary condition on the SF phase is imposed by the surrounding MI, thus the size of the region I is 
determined by the strength of the trap and can be controlled experimentally.    
In Fig.~\ref{dendifw} we plot the same density profile with the same parameters but different values of $w$. 
The chemical potential was fixed at $\mu=1.8$ as an initial condition. 
As $w$ increases, the trap potential increases and the total number of bosons required to satisfy the 
condition $\mu=1.8$ decreases. As a consequence, the system empties at the edges where the trap potential 
is stronger.  
A good way to test the validity of the local density approximation is to calculate the variance of the 
local density or local compressibility 
\be
\Delta n_i^2 = \langle n_i^2 \rangle - \langle n_i \rangle ^2 ,  
\ee
and plot it versus the local chemical potential. 
Since the CDW and the MI phases are incompressible, this quantity shows minima in the 
regions where those phases appear. Such behavior was reported in Refs.~\cite{batrouni,sara} 
for the $V=0$ case where a minimum was found precisely in the MI regions. In those regions, 
the local chemical potential $\mu$, varied 
within a range which was in agreement with the values of $\mu$ where the MI phase 
appears in the homogeneous system. This validates the use of the local density 
approximation for the $V=0$ case.      
\begin{figure}
\centering
\epsfysize=7.0truecm
\epsffile{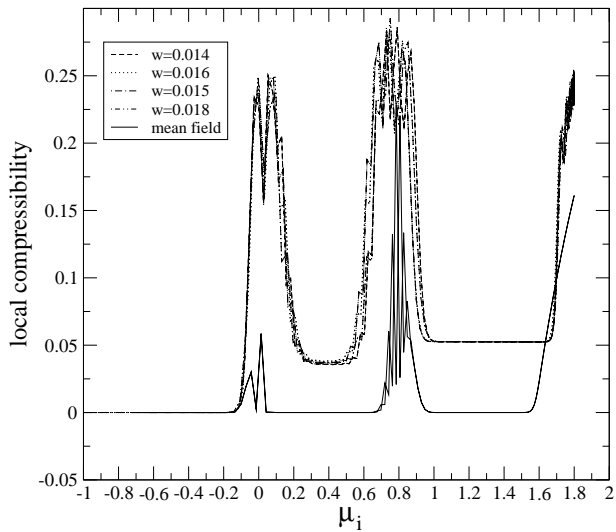}
\caption{The local compressibility vs the local chemical potential for $t=0.05$, $V=0.4$, and different values of $w$.}
\label{vardifw}
\end{figure}
In Fig.~\ref{vardifw}, we show our result for the local compressibility for the case $V=0.4$ and $t=0.05$. From left to 
right, the first minimum corresponds to the CDW phase and the second to the Mott-insulator phase. Each curve corresponds to a 
different value of $w$ and from the small differences among them we can estimate an error of $\Delta \mu=0.05$ in the 
determination of the plateau widths for the present parameter range. 
The curve in black shows the result from the mean field approximation (MFA) which, for 
this value of $t=0.05$, is acceptable although the CDW plateau is clearly much wider compared with the DMRG results. We will 
see that the MFA works more poorly as we go to larger values of $t$.         
In Fig.~\ref{vardift}, we plot the local compressibility for $V=0.4$, $w=0.015$ and different values of $t$. Within the 
local density approximation, the edges of the 
plateaus determine the limits of the regions where the CDW and the MI phases appear \cite{sara}.
\begin{figure}
\centering
\epsfig{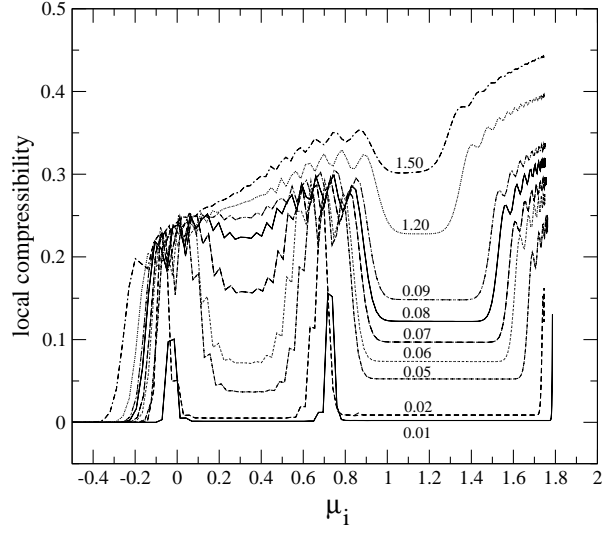}
\caption{The local compressibility vs the local chemical potential for different values of $t$, with $V=0.4$ and 
$w=0.015$.}
\label{vardift}
\end{figure}
The values of the local chemical potential obtained in this way are in good agreement with those of the phase diagram obtained in 
\cite{kuhner} for the homogeneous system. The chemical potential was initially 
set at $\mu=1.8$ in the MF ansatz, but was 
recalculated as $\mu=dE_{gs}/dN_{bos}$ using the DMRG results in order to locate exactly the curves in Fig.~\ref{vardift}. This 
calculation is shown in Fig.~\ref{enermuc} where the ground state energy $E_{gs}$ is plotted versus the total number of bosons 
$N_{bos}$ and the chemical potential is obtained from the slopes. We can see that the recalculated chemical potentials differ 
very little from the initial value $1.8$. The agreement of the results shown in Fig.~\ref{vardift} and the phase diagram 
found  in \cite{kuhner} for the homogeneous system, validates the use of the local density approximation in the inhomogeneous 
system as a way of obtaining the phase diagram of the homogeneous one. 
\begin{figure}
\centering
\epsfysize=7.0truecm
\epsffile{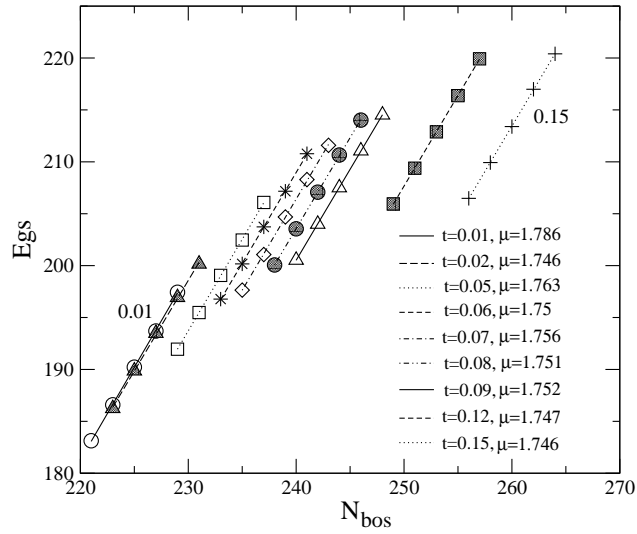}
\caption{Ground state energy vs the total number of bosons for different values of $t$, and $U=1$, $V=0.4$, and $w=0.015$.}\label{enermuc}
\end{figure}
In Fig.~\ref{pdia3} we incorporate the results obtained from 
the plateaus to the phase diagram shown in Fig.~\ref{ebhdia}. The error bars (approx $\pm 0.05$) are estimated from the variations 
we found when we changed $w$ and also from the error in the determination of the chemical potential.  
\begin{figure}
\centering
\epsfysize=7.0truecm
\epsffile{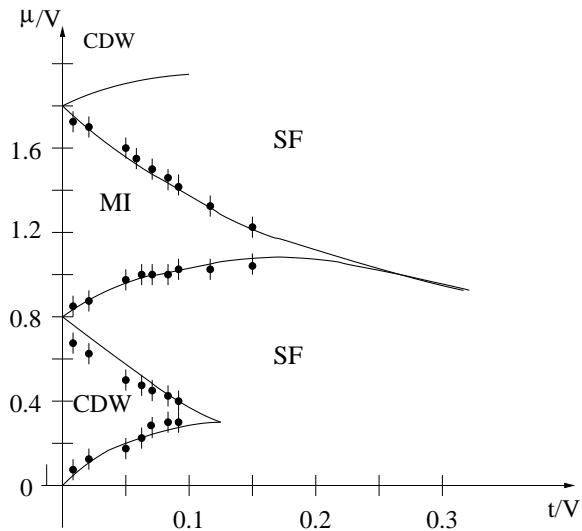}
\caption{Phase diagram of the EBH model with $V=0.4$. The symbols with error 
bars represent the results from the present study with a confining trap, 
and the lines were obtained by  K\"uhner {\it et al.} for a homogeneous 
system.}
\label{pdia3}
\end{figure}
As we mentioned before, the MFA doesn't work well as we go to larger values of $t$. In Fig.~\ref{vart015} we compare   
the local compressibility results obtained with the MFA and the DMRG for $t=0.15$. We can observe the improvement achieved with 
the DMRG method. At this value of $t$ we expect to observe only one plateau corresponding to the MI phase. The MFA curve shows 
two local minima; the first, corresponding to the CDW phase, shouldn't appear and the second, corresponding to the MI, should be 
wider. On the other hand, the DMRG curve shows only one plateau corresponding to the MI phase which agrees with previous 
results \cite{kuhner}.    
\begin{figure}
\centering
\epsfig{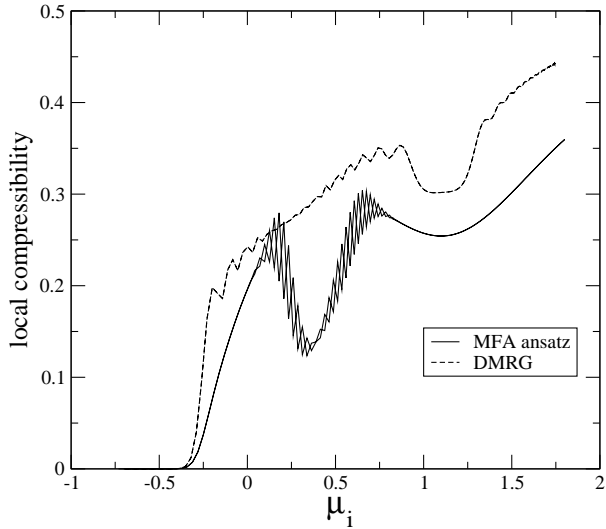}
\caption{Comparison between the local compressibilities obtained with the MFA and the DMRG for $t=0.15$, and $V=0.4$ and 
$w=0.015$.}
\label{vart015}
\end{figure} 
Other quantities which are interesting to study are the correlation functions. In particular, we computed the hopping correlation 
functions defined by $\Gamma(r-r_0)=\langle b^{\dagger}_r \, b_{r_0} \rangle$. The site labeled $r_0$ was always located at least 
20 lattice spacings from the edge of the chain to avoid boundary effects and $\Gamma(r-r_0)$ was calculated varying $r$ along the 
system in the last sweep of the DMRG. In Fig.~\ref{bcorrt005} we show these correlations where each curve corresponds to a 
different starting site $r_0$.  
Looking at this figure, we can see that there are dramatic changes in the derivative of the correlations located around the sites 
where the phases meet 
(cf.\ Fig.~\ref{bosden1}). 
If we relate the derivative of the correlations to the inverse of the correlation length we find the consistent 
result that the slopes reduce considerably in the SF regions. This is consistent because 
in the homogeneous infinite system we should expect an algebraic decay of the correlations in the SF phase (i.e. an  
infinite correlation length) and an exponential decay in the CDW and MI phases (i.e. a finite correlation 
length). Since we have an inhomogeneous system with phase separation, we cannot expect 
the long-range behavior to be strictly exponential or algebraic. 
Even when the two points used to evaluate the correlations are inside a sector 
corresponding to a given phase, there will be finite-size effects as well as additional effects coming from the phase boundaries. 
Also, dealing with a finite system means that the energy spectrum will have a gap, which implies that all correlations will eventually decay 
exponentially so that the correlation length will always be finite. 
However, even in this finite system the characteristics of the phases are 
clearly visible, with a drastic change of the correlations between the SF,
CDW and MI regions, so we conclude that the correlation analysis gives us 
another way to analyze the phase diagram in addition to the local compressibility.      
\begin{figure}
\centering
\epsfysize=7.0truecm
\epsffile{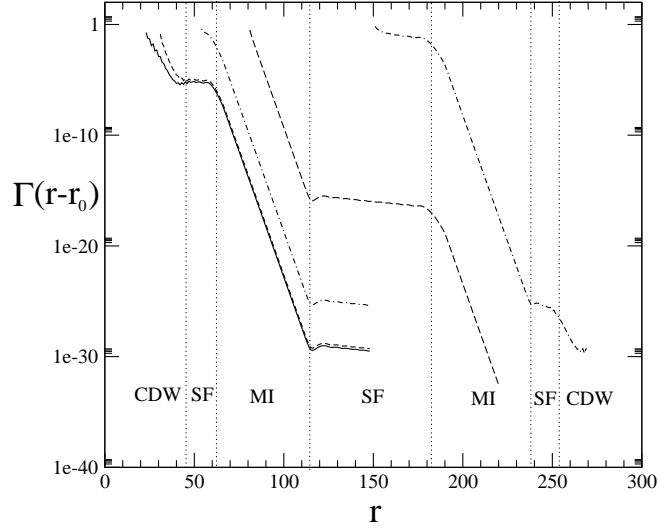}
\caption{Logarithm of the correlation functions $\Gamma (r-r_0)$ vs $r$ for $t=0.05$, and $V=0.4$ and $w=0.015$. 
The different curves correspond to different starting sites $r_0$ to compute the correlations.}
\label{bcorrt005}
\end{figure} 
In Fig.~\ref{bcorrdift}, the correlations are analyzed for different values of $t$. As we move toward larger values of 
$t$ the slopes in the CDW and MI regions decrease and the SF regions become wider. This is consistent with the fact that 
the MI and CDW phases disappear at large $t/U$. Only the SF phase, characterized by an infinite correlation length, is 
present at large $t/U$.   
\begin{figure}
\centering
\epsfysize=7.0truecm
\epsffile{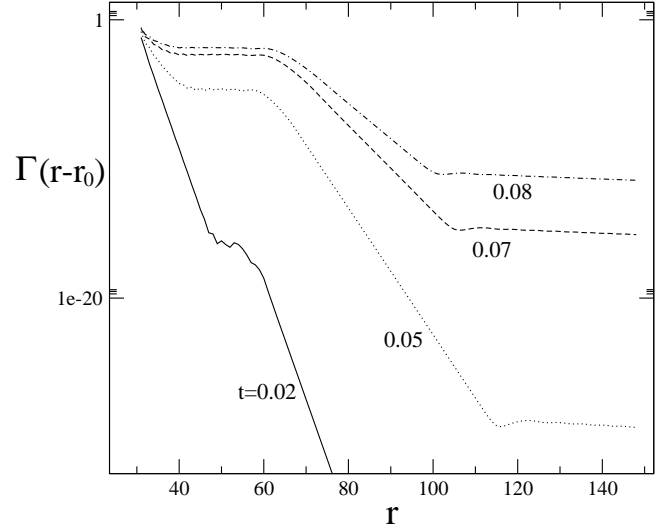}
\caption{Logarithm of the correlation functions for different values of $t$, and $V=0.4$ and $w=0.015$. As we move 
towards larger values of $t$ the slopes in the CDW and MI regions decrease and the SF regions become wider.}
\label{bcorrdift}
\end{figure}   
\section{Conclusions} \label{concsec}
We studied the extended Bose-Hubbard model in a quadratic trap potential using a DMRG method and computed 
the boson density profiles, the local compressibility and the hopping correlation functions. 
Due to the trap, the system becomes inhomogeneous and different phases of the homogeneous system coexist in spacially 
separated regions \cite{hebert}. In this work we showed how the density-matrix renormalization group method can be 
applied to such systems. The use of a finite-size DMRG with a mean field ansatz proved to be successful in addressing 
the EBH model confined in a trap. 

We observed the persistence of the CDW phase in a trapped geometry \cite{hebert} and tested the validity of the local 
density approximation. The local compressibility results showed minima that extended along the regions where the CDW 
and MI phases appeared, in agreement with the fact that those phases are incompressible. The plateaus obtained in this 
way for the CDW and MI phases allowed us to determine the phase separation lines of the phase diagram of the homogeneous 
EBH model with $V=0.4$ which is in agreement with previous results \cite{kuhner}. Therefore, our study confirms the 
validity of the local density approximation in this system. 

The correlation function results were also consistent with all the previous results. The abrupt change observed in the 
slopes at the boundaries between the coexisting phases gives an alternative way to study the phase diagram of this 
model.     
\section*{Acknowledgments}
L.U would like to thank Martin Kruczenski for discussions.
This work was supported by the Swedish Research Council.

\end{document}